\documentclass{elsart}

\usepackage{latexsym}
\usepackage{amssymb}
\usepackage{amsmath}
\usepackage{verbatim}
\usepackage{moreverb}
\usepackage{array}
\usepackage{color}
\usepackage[dvips]{graphicx}
\usepackage{epsfig}

\newcommand{\K}[1]{\ensuremath{\mathbb{K}^{#1}}}

\newcommand{\R}[1]{\ensuremath{\mathbb{R}^{#1}}}
\newcommand{\Rr}[0]{\ensuremath{\mathbb{R}}}
\newcommand{\Z}[1]{\ensuremath{\mathbb{Z}^{#1}}}
\newcommand{\Zz}[0]{\ensuremath{\mathbb{Z}}}
\newcommand{\CD}[1]{\ensuremath{\mathbb{C}^{#1}}}

\newcommand{\Cl}[1]{\ensuremath{\mathrm{Cl}(#1)}}
\newcommand{\Op}[1]{\ensuremath{\mathrm{Op}(#1)}}

\newcommand{\Coord}[1]{\ensuremath{x^{#1}}}
\newcommand{\GCoord}[2]{\ensuremath{{#1}^{#2}}}
\newcommand{\Coordmax}[1]{\ensuremath{M^{#1}}}
\newcommand{\KCoord}[1]{\ensuremath{x_K^{#1}}}

\newcommand{\KCode}[1]{\ensuremath{\mathtt{code}({#1})}}

\newcommand{\topo}[1]{\ensuremath{\mathtt{topo}{(#1)}}}
\newcommand{\orth}[1]{\ensuremath{\perp\!({#1})}} 
\def\cp{\ensuremath{p}}
\def\cq{\ensuremath{q}}
\def\cc{\ensuremath{c}}
\def\cb{\ensuremath{b}}
\def\ci{\ensuremath{i}}
\def\lbdry{\ensuremath{\Delta}}
\def\ubdry{\ensuremath{\nabla}}

\newcommand{\SpelCode}[7]{%
\setlength{\arrayrulewidth}{0.2pt}%
\addtolength{\extrarowheight}{-1mm}%
\ensuremath{%
\begin{array}{|c|c|c|c|c|c|c|}\hline%
#1 & #2 & #3 & #4 & #5 & #6 & #7 \\%
\hline%
\end{array}}}

\newcommand{\USpelCode}[6]{%
\setlength{\arrayrulewidth}{0.2pt}%
\addtolength{\extrarowheight}{-1mm}%
\ensuremath{%
\begin{array}{|c|c|c|c|c|c|}\hline%
#1 & #2 & #3 & #4 & #5 & #6 \\%
\hline%
\end{array}}}

\newenvironment{myarray}[1]{%
\setlength{\arrayrulewidth}{0.2pt}%
\addtolength{\extrarowheight}{-0.2mm}%
\ensuremath{\begin{array}{#1}}}%
{\end{array}}

\newcommand{\SpelCodeD}[3]{%
\SpelCode{#1}{#2}{\Coord{n-1}}{\ldots}{#3}{\ldots}{\Coord{0}}}

\newcommand{\USpelCodeD}[2]{%
\USpelCode{#1}{\Coord{n-1}}{\ldots}{#2}{\ldots}{\Coord{0}}}

\newcommand{\eg}{e.g.}

{\begin{list}{-~}{\setlength{\labelsep}{0pt}\setlength{\leftmargin}{1cm}%
\setlength{\labelwidth}{0pt}\setlength{\listparindent}{0pt}%
\setlength{\parsep}{0pt}}}%
{\end{list}}

\newcommand{\keyw}[1]{{\bf #1}}
\newcommand{\comm}[1]{{\em // #1}}
\newcommand{\mquad}{~~}
\newenvironment{algo}{\begin{tabbing} \mquad \=\mquad \=\mquad \=\mquad \=\mquad \=\mquad \=\mquad \=\mquad \kill}{\end{tabbing}}

\begin{document}

\begin{frontmatter}

\title{Coding cells of digital spaces: a framework to write
generic digital topology algorithms}

\author[Bordeaux]{Jacques-Olivier Lachaud}
\address[Bordeaux]{LaBRI, Univ. Bordeaux 1, 351 cours de la Lib\'{e}ration, 33405 Talence, France}

\begin{abstract}
This paper proposes a concise coding of the cells of $n$-dimensional
finite regular grids. It induces a simple, generic and efficient
framework for implementing classical digital topology data structures
and algorithms. Discrete subsets of multidimensional images ({\eg}
regions, digital surfaces, cubical cell complexes) have then a common
and compact representation. Moreover, algorithms have a
straightforward and efficient implementation, which is independent
from the dimension or sizes of digital images. We illustrate that
point with generic hypersurface boundary extraction algorithms by
scanning or tracking. This framework has been implemented and basic
operations as well as the presented applications have been benchmarked.
\end{abstract}

\end{frontmatter}

\section{Introduction}

Many applications in the image analysis field need to represent
and manipulate discrete subsets of digital spaces. As the image data
become larger, the data structures required to represent these sets
should be as compact as possible. Moreover, algorithms designed on
these structures should be not only efficient theoretically, but also
efficient in practice. Algorithms defined formally should have a
straightforward implementation. Last but not least, 3D and 4D image
datasets are now more and more common. It becomes necessary to have a
unified framework for programming applications dealing with
$n$-dimensional data. By this way, algorithms are both generically
defined and implemented.

There exist several approaches to define the topology of
multidimensional regular digital spaces: (1) adjacency graphs as
pioneered by Rosenfeld, (2) oriented graphs as proposed
by Herman \cite{Herman92}, (3) cellular
complexes as proposed by Kovalevsky \cite{Kovalevsky89}, or
equivalently Khalimsky's spaces \cite{Khalimsky90a} and interpixel
representations. This paper deals mostly with the third approach,
although our framework can express either approaches
(the two first approaches manipulate a restricted set of the elements
defined in the spaces of the third approach).

The cellular decomposition of the Euclidean $n$-dimensional space
\R{n} into a regular grid forms a cellular complex \CD{n}. This
structure has been introduced in digital topology by
Kovalevsky~\cite{Kovalevsky89} for 2D and 3D applications. It has been
shown that the topology induced on \CD{n} is equivalent to a digital
topology \K{n}, sometimes called {\em Khalimsky} topology
\cite{Khalimsky90a,Kong91a}.  Many authors have explored the
theoretical properties of the space \CD{n} (or equivalently \K{n})
\cite{Khalimsky90a,Kong91a,Kovalevsky89} 
applications
\cite{YBertrand99,Braquelaire99a,Kovalevsky00,Kovalevsky01,Metz94}.
These works show that the definition of consistent high level data
structure over images relies on a low-level representation which is
the regular cellular decomposition of the image support. It is thus
critical to represent efficiently arbitrary cells of \CD{n}, small
subsets of \CD{n} and specific subsets of \CD{n} ({\eg} complexes,
digital surfaces). However, the litterature does not reflect this
observation.  Indeed, spels are often coded with an array of coordinates,
surfels as pairs of adjacent spels or a spel with a direction, other cells are
generally implicitly represented.  Consequently, storing elements or
subsets of \CD{n} is cumbersome; algorithms are frequently rewritten
at the implementation stage to avoid any reference to non elementary
kinds of cells. 

In this paper, we choose another approach, which is first to show how
to represent an arbitrary cell of \CD{n} with a binary cell code and
secondly to design data structures over this representation.  Because
of the regularity of \CD{n}, each cell code holds all the information
on the cell: the cell topology (dimension, open or closed along a
coordinate, adjacent and incident cells) and geometry (coordinates in
\Z{n}, centroid, normal and tangent vectors) can be computed from the
code without any other information.  The proposed framework is suited
both to formal representation and proofs and to straightforward
implementation in a programming language. The paper is organized as
follows: (i) coding of cells and implementation of low-level digital
topology definitions, (ii) definition of data structures for subsets
of \CD{n} ({\eg}, digital surfaces, complexes), (iii) application to
digital boundaries extraction in multidimensional images.  We
emphasize that all operations and algorithms have the same definitions
and implementation whatever the dimension of the space. All
experiments and benchmarks presented in this paper were made on a PC
with a Celeron 500Mhz processor, 128Mb of memory, 128Kb of cache
(which is a basic workstation). The proposed framework was implemented
in C++. Due to limited space, the reader is referred to
\cite{Lachaud02a} for more details.

\section{Coding cells of digital spaces \CD{n}}

\subsection{Cellular decomposition \CD{n}; binary coding of unoriented cells}

We denote by \CD{n} the set of parts of the $n$-dimensional Euclidean
space \R{n} such that $c \in \CD{n}$ is equivalent to $c=I_1 \times
\ldots \times I_n$ where $I_i$ is a subset of {\Rr} of the form $\{
z_i \}$ or $\rbrack z_i, z_{i} + 1 \lbrack$ with $z_i \in {\Zz}$. The
complex \CD{n} is a partition of \R{n}.  We call {\em $k$-cell} an
element $c \in \CD{n}$ such that $c$ has $k$ $I_i$ of the form
$\rbrack z_i, z_{i}+1 \lbrack$ (and therefore $(n-k)$ $I_i$ of the
form $\{z_i\}$). The {\em dimension} of $c$ is $k$.
The {\em closure $\Cl{c}$} of a cell $c$ is the set of cells $c'$ of
$\CD{n}$ which have the following form: (i) on coordinate where $c$ is
a point $\{z_i\}$, the cell $c'$ must also be the same point, (ii) on
coordinate where $c$ is an open segment $\rbrack z_i, z_{i}+1
\lbrack$, $c'$ can be either the same open segment or the point
$\{z_i\}$ or the point $\{z_{i}+1\}$.
The {\em open star $\Op{c}$} of a cell $c$ is the set of cells $c'$ of
$\CD{n}$ such that $c' \in \Op{c} \Leftrightarrow c \in \Cl{c'}$.  The
{\em bounding relation $<$} between two cells $c$ and $c'$ is then
defined as $c < c'$ iff $c \in \Cl{c'} \setminus c'$. With these
definitions, the set $\CD{n}$ equipped with the dimension mapping and
the bounding relation is a cellular complex. A {\em cubical cell
complex} $K$ is then defined as any set of cells in a finite
image. The {\em dimension} of $K$ is the maximum of the dimensions of
its cells. Open stars and closure of cells in a complex $K$ are defined
naturally.

It is clear that the elements of $\CD{n}$ represent low-level elements
of $n$-dimensional digital images: the {\em spels} (pixels in 2D and
voxels in 3D) are the $n$-cells, the {\em (unoriented) surfels} (a
pair of adjacent spels) are the $n-1$-cells, the vertices of the spels
and of the surfels, or {\em pointels}, are the $0$-cells.  An {\em
object} is then a set of $n$-cells, a {\em digital surface} is a set
of $n-1$-cells (oriented or not, see
Section~\ref{sec:coding-oriented-cells}), a curve is a set of
connected $1$-cells. Therefore, all classical subsets of digital
spaces have a natural definition as specific subsets of \CD{n}.
From now on, we will assume that we are working in a finite
$n$-dimensional image forming a parallelepiped in \Z{n}. We denote by
$\Coordmax{i}$ the inclusive upper bound for the $i$-th coordinate of
any spel. All coordinates have 0 as lower bound.


As shown by Kong \etal~\cite{Kong91a}, the topology of \CD{n} is
equivalent to the topology of the Khalimsky digital space \K{n}, which
is the cartesian product of $n$ connected ordered topological spaces
(COTS). A COTS can be seen as a set of ordered discrete points, like
\Zz, whose topology alternates closed points and open points. If we
define even points of {\Zz} as closed and odd points of {\Zz} as open,
each point of \K{n} 
is then identified by its $n$ integer coordinates, whose parities
define its topological properties.

\begin{sloppypar}
Consequently, any $k$-cell $c$ of \CD{n} has exactly one corresponding
point in \K{n} with coordinates $(\KCoord{0}, \ldots,
\KCoord{n-1})$. We propose to code {\cc} as one binary word $\KCode{c} =
\USpelCodeD{\alpha}{\Coord{i}}$, called the {\em unsigned code} of
{\cc}, as follows:
\end{sloppypar}
\begin{itemize}

  \item The $i$-th coordinate \KCoord{i} is coded by its
  binary decomposition after a rightshift ($\Coord{i} = \KCoord{i}
  \mathrm{~div~} 2$). We say that $\Coord{i}$ is the {\em
  $i$-th digital coordinate} of $c$.

  \item All coordinates are packed as one binary word (from \Coord{n-1}
  to \Coord{0}). Every coordinate is allocated a fixed number of bits
  $N_i$ given by $N_i = \log_2(\Coordmax{i})+1$.

  \item The parity of all coordinates are also packed as an $n$-bits
  word $\alpha$ with $\alpha = \sum_i (\KCoord{i} \mathrm{~mod~}
  2) 2^i$. $\alpha$ is called the {\em topology} of $c$.

\end{itemize}

Spels have a topology word composed of 1's, whereas pointels have a
topology word made of 0's. Surfels have one 0 and $n-1$ 1's in their
topology word. The coordinate where a surfel {\cc} has a 0 in its
topology word is called the coordinate {\em orthogonal} to the surfel
{\cc} and is denoted by \orth{\cc}.  This coding implies that any cell
of finite digital images can be coded as an integer number with fixed
size. Any register of a processor may thus store a cell if the image is not
too big.\footnote{32 bits are sufficient to code every cell of a $1024
\times 1024 \times 512$ 3D image, which is more than enough for
current biomedical applications.}  We define the adjacency between
cells independently of the cell topology.
\begin{defn}
Two cells {\cp} and {\cq} with $\topo{\cp} = \topo{\cq}$
are {\em $l$-adjacent} if their respective coordinates \GCoord{p}{i}
and \GCoord{q}{i} differ by at most $1$ and if the infinite norm of
the vector $(\GCoord{p}{n-1}-\GCoord{q}{n-1}, \ldots,
\GCoord{p}{0}-\GCoord{q}{0})$ is no more than $l$.
\end{defn}

The $1$-adjacency thus defines the 4-adjacency (resp. 6-adjacency) on
pixels in 2D (resp. on voxels in 3D) and the $2$-adjacency defines the
8-adjacency (resp. 18-adjacency) on pixels in 2D (resp. on voxels in
3D).
We define the incidence relation as below. The proposition that
follows shows that all the topological structure of \CD{n} can be
obtained from the incidence relation.
\begin{defn}
\begin{sloppypar}
Let ${\cc}=\USpelCodeD{\alpha}{\Coord{i}}$ be a cell and $i$ any
coordinate. Let $\beta=\alpha \mathrm{~xor~} 2^i$. If the $i$-th bit
of $\alpha$ is set to 1, the cell {\cc} has two {\em low 1-incident
cells} along coordinate $i$ coded by $\USpelCodeD{\beta}{\Coord{i}}$
and $\USpelCodeD{\beta}{\Coord{i}+1}$.  Otherwise, if the $i$-th bit
of $\alpha$ is set to 0, the cell {\cc} has two {\em up 1-incident
cells} along coordinate {\ci} coded by
$\USpelCodeD{\beta}{\Coord{i}-1}$ and $\USpelCodeD{\beta}{\Coord{i}}$.
A cell {\cp} is {\em low incident} (resp. {\em up incident}) to a cell
{\cq} if there is a sequence of cells ${c_0}=\cp, {c_1},
\ldots, {c_k}=\cq$ such that $\forall j$, ${c_j}$ is low
1-incident (resp. up 1-incident) to ${c_{j+1}}$.
\end{sloppypar}
\end{defn}

\begin{prop}
The set of cells low incident to a cell {\cc} is equal to $\Cl{\cc}
\setminus \cc$. The set of cells up incident to {\cc} is equal to
$\Op{\cc} \setminus \cc$.
\end{prop}

\begin{figure}

\begin{tabular}{cccccccccc}
nb ops & & topo, &
& set & & is &  & is \\[-2.5mm]
required & code & coord & {\tt ==}
& coord & adj. & $l$-adj.? & inc. &
$l$-inc.? \\ \hline
bits ops     & 0   & 1 & 0 & 2 & 0 & $\leq 2n$ & 1        & $\leq 3$ \\[-1.5mm]
shifts       & $n$ & 1 & 0 & 1 & 0 & 0         & 0        & $\leq 6$ \\[-1.5mm]
integer ops  & $n$ & 0 & 1 & 0 & 1 & $\leq 2n$ & $\leq 1$ & $\leq l+4$ \\[-1.5mm]
lut access   & $n$ & 1 & 0 & 2 & 1 & $\leq n$  & $\leq 2$ & $\leq l+2$ \\[-1.5mm]
cond. tests  & 0   & 0 & 0 & 0 & 1 & $\leq 2n$ & 1        & $\leq 3l+1$ \\[-1.5mm]
\end{tabular}

\caption{Number of elementary operations needed to perform the
following tasks: (i) coding a vector of $n$ Khalimsky coordinates as a
cell, (ii) getting the topology or one coordinate of a cell, (iii) comparing if two cells are identical, (iv)
setting the coordinate of a cell, (v) computing a 1-adjacent cell,
(vi) checking if two cells are $l$-adjacent, (vii) computing a
1-incident cell, (viii) checking if two cells are $l$-incident.}
\label{fig:elementary-ops}
\end{figure}

Figure~\ref{fig:elementary-ops} summarizes the number of elementary
operations necessary to execute basic cell operations. Their
implementation is fully generic. We have benchmarked these operations
and the results show that cell codes compete with statically defined
structures (e.g. fixed size arrays) and are much faster than
dynamically allocated structures, classically used for generic
programming.


\subsection{Oriented cells, boundary operators, bels, boundary of an object}
\label{sec:coding-oriented-cells}

In some applications, it is convenient to orient the cells (as
positive or negative). For instance, digital surfaces as proposed by
Herman and Udupa are composed of oriented pairs of voxels: one voxel
is in the interior of the surface, the other in the
exterior. Orienting a surfel means in this case to define where are
the interior and exterior voxels 1-up-incident to the surfel. Digital
surface tracking algorithms rely on this orientation for a consistent
output. Classical combinatorial topology
associates an orientation to each cell of a cellular complex. Oriented
cells are then useful to implement boundary operators over complexes
and to compute topological invariants.

We therefore define the {\em signed code} of a cell
{\cc} with orientation bit $s$ (0 is positive, 1 is negative) by
adding the bit $s$ between the topology $\alpha$ of {\cc} and its
digital coordinates $\Coord{i}$ as follows:
\SpelCodeD{\alpha}{s}{\Coord{i}}. The {\em opposite cell $-{\cc}$} of
{\cc} is the same cell as {\cc} but with opposite sign.  
Boundary operators, which can be seen informally as an oriented
version of incidence, are essential in combinatorial topology~: for
instance, they define the topology of polyhedral complexes. 
We have now to ``orient'' the incidence relation.

\begin{defn}
\begin{sloppypar}
Let ${\cc} = \SpelCodeD{i_k \ldots i_j \ldots i_0}{s}{\Coord{i_j}}$ be
any cell with topology bits set to 1 on the coordinates $i_k, \ldots,
i_j, \dots, i_0$, $n-1 \ge i_k > \ldots > i_j > \dots > i_0 \ge 0$ and
the others bits set to 0. The symbol $\hat{i}_j$ means that the bit
$i_j$ is set to 0. 
Let $\tau = (-1)^{(k-j)}$.
%
The set $\lbdry_{i_j} \cc$ composed of the two oppositely signed cells
$\tau \SpelCodeD{i_k \ldots \hat{i}_j \ldots i_0}{s}{\Coord{i_j}}$ and
$- \tau \SpelCodeD{i_k \ldots \hat{i}_j \ldots
i_0}{s}{\Coord{i_j}+1}$, is called the {\em lower boundary of the cell
{\cc} along coordinate $i_j$}.  The {\em lower boundary} $\lbdry{\cc}$
of {\cc} is then the set of cells $\cup_{l=0,\ldots,k}
\lbdry_{i_l}{\cc}$. 
\end{sloppypar}
\end{defn}


The lower boundary of {\cc} thus corresponds to the set of cells
1-low-incident to {\cc} with specific orientations. The {\em upper
boundary} $\ubdry$ of a cell is defined symmetrically (the upper
boundary is taken on topology bits set to 0). It can be shown that
this definition of boundary operators induces that any cubical cell
complex is a polyhedral complex. Although this is outside the scope of
this paper, boundary operators along with {\em chains} are used to
define the (co)homology groups of complexes, which are well known
topological invariants. This topological tool is readily applicable to
cubical cell complexes.

In the remainder of the paper, the set $O$ is an object of the image
$I$ with an empty intersection with the border of $I$. Assume that all
spels of $O$ are oriented positively. We merge the sets $\lbdry{\cp}$
with ${\cp} \in O$ with the rule that two identical cells except for
their orientation cancel each other.  The resulting set of oriented
surfels is called the {\em boundary of $O$}, denoted by $\partial
O$. It is a digital surface, whose elements are called {\em bels} of
$O$. The following result states that the boundary of $O$ is indeed
the digital surface separating the spels of $O$ from the spels of the
complement of $O$.

\begin{prop}
Let {\cc} be a bel of $\partial O$. Then $\ubdry{\cc}$ contains two
spels: one positively oriented and belonging to $O$, the other
negatively oriented and not belonging to $O$. Any path of 1-adjacent
spels from an element of $O$ to an element not in $O$ crosses
$\partial O$.
\end{prop}

One way to compute the digital surface bordering a set of spels $O$ is
by applying the lower boundary operator on each element of $O$ and
removing oppositely oriented identical cells. The time complexity of
this algorithm is thus linear with the number of spels of $O$.

\subsection{Followers of surfel, bel adjacency, digital surface tracking}
\label{sec:bel-adjacency}

The bel adjacency defines the connectedness relations between bels
bounding an object. It has two nice consequences: (i) the boundary of
an object can be extracted by tracking the bels throughout their bel
adjacencies~\cite{Artzy81,Gordon89}; (ii) sets of surfels can be
considered as classical Euclidean surfaces, where one can move on the
surface in different orthogonal directions (2 in 3D). The second
reason is essential for defining the geometry of digital surfaces
\cite{Lachaud02a}. 
We present here a definition of bel adjacencies that is essentially
equivalent to the definition of \cite{Herman98b}, but easier to
implement in our framework.  We start by defining which surfels are
potentially adjacent to a given bel with the notion of follower. We
then define two kinds of bel adjacency for each pair of coordinates.

\begin{defn}
We say that an $r$-cell {\cq} is a {\em direct follower} of an
$r$-cell {\cp}, $\cp \neq \pm \cq$, if $\lbdry{\cp}$ and $\lbdry{\cq}$
have a common $r-1$-cell,
called the {\em direct link} from {\cp} to {\cq}, 
such that this cell is positively oriented in $\lbdry{\cp}$ and
negatively oriented in $\lbdry{\cq}$. The cell {\cp} is then an {\em
indirect follower} of {\cq}.
\end{defn}

\begin{figure}
\begin{center}
\input{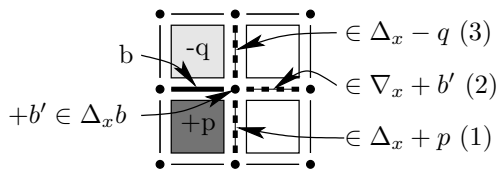}
\end{center}
\caption{Direct followers of a surfel {\cb} along coordinate $x$.}
\label{fig:bel-adjacency-illustration}
\end{figure}

It is easy to check that any surfel has 3 direct followers and 3
indirect followers along all coordinates except the one orthogonal to
the surfel. We order the followers consistently for digital surface
tracking (see Figure~\ref{fig:bel-adjacency-illustration}).
\begin{defn}
Let {\cb} be an oriented $n-1$-cell, such that
$\ubdry{\cb}=\{+{\cp},-{\cq}\}$. Let $j$ be a coordinate with $j \neq
\orth{\cp}$. The three direct followers of {\cp} along $j$ are ordered
as follows: (1) the first direct follower belongs to $\lbdry_j
+{\cp}$, (2) the second direct follower belongs to $\ubdry_j +{\cb'}$
with $+{\cb'}$ direct link in $\lbdry_j {\cb}$, (3) the third direct
follower belongs to $\lbdry_j -{\cq}$.
\end{defn}

Intuitively, when tracking a digital surface, you have 3 different
possibilities for a move along a given coordinate. This is true for
arbitrary dimension. The following definition shows which one to
choose at each step .
It is in agreement with the definitions of bel adjacencies proposed by
Udupa \cite{Udupa94}.
\begin{defn}
Let {\cb} be a bel of $\partial O$, such that
$\ubdry{\cb}=\{+{\cp},-{\cq}\}$ (thus ${\cp} \in O$ and ${\cq} \not\in
O$). For any coordinate $j \neq \orth{\cb}$, the bel {\cb}
has one {\em interior direct adjacent bel} (resp. {\em exterior direct
adjacent bel}) which is the first (resp. last) of the three ordered
direct followers of {\cb} along coordinate $j$ that is a bel of
$O$. The bel adjacency is the symmetric closure of the direct bel
adjacency.
\end{defn}
In 3D, the interior (resp. exterior) bel adjacency along all
coordinates induces the classical (6,18) bel-adjacency (resp. (18,6)
bel-adjacency). Interior and exterior bel adjacencies can be mixed for
different coordinate pairs. This might be useful in an application
where the image data are not isotropic ({\eg}, some CT scan images,
confocal microscopy). Computing the bel adjacent to a given one is very
fast since it required \cite{Lachaud02a}: $\leq 11$ binary or integer
operations, $\leq 3$ shifts, $\leq 14$ lut accesses, $\leq 9$
conditional tests, and $1$ or $2$ ``is in set'' operations. The next
section will show that the ``is in set'' operation can be done in four
elementary operations.

The following theorem, which comes from the fact that cubical cell
complexes are polyhedral complexes, is interesting to speed up digital
surface tracking algorithm: as its corollary, tracking only direct
adjacent bels is sufficient to extract the whole digital surface
component that contains the seed bel. It is more complex to show that
bel components correspond to interior and exterior components of spels
(see \cite{Lachaud00a,Udupa94} where this is proven for some bel
adjacency relations).
\begin{thm}
The transitive closure of the direct bel adjacency from a bel {\cb} of
$\partial O$ coincides with the transitive closure of the bel
adjacency from {\cb}.
\end{thm}

\section{Data structures built over cells}
\label{sec:data-structures}


Since any kind of cell is coded as an integer number, data structures
coding sets of cells are easily derived from standard data types. Table~\ref{tbl:set-data-structures} displays the traditional set data structures, their
implementation in C++ as template classes, the time complexities of
some operations, and the memory cost.\footnote{A cell is stored in a
32-bits word. We suppose that 12 bytes are necessary to store
information about one dynamically allocated memory area ({\eg} holds
for Linux)} 

\begin{table}
\caption{This table shows some properties of classical set data
structures. The symbol $+$ indicates that it is only amortized time
complexity.}
\label{tbl:set-data-structures}
\begin{center}
\begin{tabular}{ccccc}
set structure &
STL class &
is in set ? &
other set ops &
memory (bytes)
 \\ \hline
dynamic array & \texttt{vector} & $O(m)$ & $O(m)+$ & $\approx 4m$ \\
linked list & \texttt{list} & $O(m)$ & $O(m)$ & $\approx 24m$ \\
RB-tree & \texttt{set} & $O(\log m)+$ & $O(\log m)+$ & $\approx 32m$ \\
hashtable & \texttt{hash\_set} & $O(1)+$ & $O(1)+$ & $\approx 4m'+20m$ \\
\end{tabular}
\end{center}
\end{table}

These data structures are adapted to sets of cells of reasonnable
size. Very small sets should be defined as
\texttt{vector}s. \texttt{list}s may be used to represent medium size
contours. Other medium size sets should be represented with
\texttt{set}s or \texttt{hash\_set}s.  If the \texttt{hash\_set} seems
rather efficient for most operations (at least from an asymptotic
point of view), it is memory costly: $28$ Mbytes are necessary to
represent a digital surface with $1,000,000$ bels (and
$m'=2m$). Moreover the memory is very fragmented and the cache is thus
not efficient. As it is shown later on digital hypersurface tracking
algorithms, amortized constant time does not mean very fast.


We present another data structure to represent a set of cells, which
exploits the properties of the cell coding. The size of the data
structure is dependent only on the size of the image. The time
complexity of all operations is then independent from the number of
cells represented.  This data structure, called the \texttt{CharSet},
is a characteristic function that assigns one bit to each cell of the
space. Since we will often manipulate sets of cells that contains
specific kinds of cells ({\eg.} a digital surface is made of surfels),
we present two ways to define this structure.

\begin{defn}
\begin{sloppypar}
A\/ $\mathtt{MinCharSet}$ is an array\/ $\mathtt{tbl}$ of $s$ bits,
where $s$ is one plus the difference between the highest possible cell
code\/ $\mathtt{MAX}$ and the smallest possible cell code\/
$\mathtt{min}$. Selecting the bit characteristic of the presence of a
given cell {\cc} is done with\/ \verb+tbl[(c-min)>>5]&(1<<(c&0x1f))+
for 32-bits words.
\end{sloppypar}
\end{defn}
\begin{defn}
\begin{sloppypar}
A\/ $\mathtt{LUTCharSet}$ is an array\/ $\mathtt{tbl}$ of $s$ bits and
a look-up table\/ $\mathtt{lut}$, where $s$ and\/ $\mathtt{lut}$ are
dependent on the set of cells (see
Table~\ref{tbl:lutcharset}). Selecting the bit characteristic of the
presence of a given cell {\cc} is done with\/
\verb!tbl[(lut[topo(c)]+sign_coords(c))>>5]&(1<<(c&0x1f))! for 32-bits
words.
\end{sloppypar}
\end{defn}

The \texttt{LUTCharSet} is more compact than the \texttt{MINCharSet}
for some sets of cells (and the higher the dimension the more it is)
but the access to the characteristic bit of a cell is a bit slower It
is now clear why the bit defining the sign of an oriented cell is
inserted between the topology and the coordinates of the cell: with
this coding, both \texttt{CharSet}s use exactly twice more memory for
sets of signed cells compared with sets of unsigned cells.

\begin{table}
\caption{This table defines the way \texttt{LUTCharSet}s store various
specific sets of cells.}
\label{tbl:lutcharset}

\begin{tabular}{m{1.7cm}m{2.3cm}m{3.6cm}m{1.7cm}m{2.3cm}}
set of cells & topologies $\alpha$ & $\mathtt{lut}(\alpha)$
& size $s$ (bits) & $256^3$ image size (Mb) \\ \hline
set of spels & 
$\begin{myarray}{|c|}\hline 1 \ldots 1 \\ \hline \end{myarray}$ & 
$\begin{myarray}{|c|c|}
\hline 
0 & 0 \ldots 0 \\ \hline 
\multicolumn{1}{c}{~}& \multicolumn{1}{c}{\sum N_i \mathrm{~bits}} \\
\end{myarray}$ 
& $2^{\sum N_i}$ & 2 
\\[2mm]
unoriented digital surface & 
$\begin{myarray}{|c|}\hline 
011 \ldots 1 \\ \hline 
101 \ldots 1 \\ \hline 
 \multicolumn{1}{c}{\ldots} \\ \hline 
1 \ldots 110 \\ \hline 
\multicolumn{1}{c}{~} \\
\end{myarray}$ & 
$\begin{myarray}{|c|c|}
\hline
0 & 0 \ldots 0 \\ \hline
1 & 0 \ldots 0 \\ \hline
\multicolumn{2}{c}{\ldots} \\ \hline
n-1 & 0 \ldots 0 \\ \hline
\multicolumn{1}{c}{~}& \multicolumn{1}{c}{\sum N_i \mathrm{~bits}} \\
\end{myarray}$ 
& $n 2^{\sum N_i}$ & 6
\\[2mm]
set of oriented $r$-cells &
$\begin{myarray}{|c|}
\hline
0 \ldots 0011 \ldots 1 \\ \hline 
0 \ldots 0101 \ldots 1 \\ \hline 
\multicolumn{1}{c}{\ldots} \\ \hline 
1 \ldots 1100 \ldots 0 \\ \hline 
\multicolumn{1}{c}{~} \\
\end{myarray}$ &
$\begin{myarray}{|c|c|}
\hline
0 & 0 \ldots 0 \\ \hline
1 & 0 \ldots 0 \\ \hline
\multicolumn{2}{c}{\ldots} \\ \hline
\binom{n}{r} - 1 & 0 \ldots 0 \\ \hline
\multicolumn{1}{c}{~} & \multicolumn{1}{c}{1+\sum N_i \mathrm{~bits}} \\
\end{myarray}$ &
$ 2\binom{n}{r} 2^{\sum N_i}$
& 12 ($r=1,2$) 
\\
\end{tabular}
\end{table}

\begin{sloppypar}
Knowing if a cell belongs to a \texttt{CharSet} or any other atomic
set operation (add/remove an element) are $O(1)$ operations.  All
global set operations (like union, intersection, difference,
complement) between \texttt{CharSet}s are implemented as standard bit
operations between arrays of binary words. Their time complexities are
linear in the size of the array. Moreover, the implementation of set
operations for any set of cells (arbitrary dimension, set of spels,
oriented digital surface, set of $r$-cells, etc) is done only once as
bit operations between arrays of binary words.  To give an idea of the
efficiency of this representation, inverting a set of spels defined in
a $512^3$ image takes $0.40$s ($134,217,728$ spels, $3$ns per spel),
difference between two sets of spels in the same image takes
$0.80$s. Furthermore, an unoriented digital surface in a $256^3$ image
can hold up to $50,331,648$ surfels for a $8$Mb memory cost (or $6$Mb
for \texttt{LUTCharSet}).
To conclude this section, unsigned sets are twice less costly to
store. They should be used when possible. For instance, digital
surfaces that are boundaries of a set of spels are always orientable
surface. Digital surface tracking can thus be done with unoriented
digital surfaces. 
\end{sloppypar}


\section{Digital boundary extraction by scanning and tracking}

We have implemented several digital hyper-surface extraction
algorithms which build the digital surface that is the boundary of a
given object $O$. Scanning algorithms examine every spel neighborhood
to detect the presence of a bel. They only require the set $O$ as
input. Digital surface tracking algorithms require an initial bel
{\cb} and a bel adjacency \texttt{A} to extract the component of the
boundary of $O$ that contains {\cb}. As described in
Section~\ref{sec:bel-adjacency}, defining the bel adjacency \texttt{A}
is deciding for each pair of coordinates whether \texttt{A} is
interior or exterior along this plane.
Figure~\ref{fig:digital-surface-tracking} shows how to write generic
digital surface tracking algorithms with our framework. The
implementation in C++ is very close to the formal specification of the
algorithm (see \cite{Herman98b}).

\begin{figure}
\begin{minipage}{0.5\linewidth}
\begin{algo}
\comm{Track (B) algorithm.} \\
\comm{$\partial O$ must be closed.} \\
CharSet \\
Space::track( CharSet O, Cell b,\\
\> \> \> \> \> \>BelAdj A ) \\
\{\>CharSet S = emptySurfelSet(); \\
\>  queue$<$Cell$>$ L; \comm{queue of bels} \\
\>  L.push( b ); \comm{starting bel} \\
\>  \keyw{while} ( ! L.empty() ) \{ \\
\>\>    Cell p = L.pop(); \comm{current bel} \\
\>\> \comm{On all coord where p open} \\
\>\>    \keyw{for} ( int j = 0; j $<$ dim(); ++j ) \\
\>\>\>      \keyw{if} (j != orthDir(p)) \{ \\
\>\>\>\> \comm{Track direct followers} \\
\>\>\>\>        Cell q = A.directAdj(O,p,j); \\
\>\>\>\>        \keyw{if} (! S.isInSet(q)) \{ \\
\>\>\>\>\>          S.add( q ); \\
\>\>\>\>\>          L.push( q ); \\
\>\>\>\> \} \\
\>\>\> \}  \\
\>\} \\
\>  \keyw{return} S; \\
\}
\end{algo}
\end{minipage}
\begin{minipage}{0.5\linewidth}
\begin{algo}
\comm{Track (C) algorithm.}  \\
\comm{$\partial O$ must be closed.} \\
CharSet  \\
Space::track( CharSet O, Cell b, \\
\> \> \> \> \> \>BelAdj A ) \\
\{\>CharSet S = emptySurfelSet(); \\
\>  queue$<$Cell$>$ L; \comm{queue of bels} \\
\>  list$<$Cell$>$ T; \comm{"tail" of bdry} \\
\>  L.push( b ); \comm{starting bel} \\
\>  T.multipleInsert( b, dim() - 1 );  \\
\>  \keyw{while} ( ! L.empty() ) \{ \\
\>\>    Cell p = L.pop(); \comm{current bel} \\
\>\>    \keyw{for} ( int j = 0; j $<$ dim(); ++j ) \\
\>\>\>      \keyw{if} ( j != orthDir( p ) ) \{ \\
\>\>\>\>        Cell q = A.directAdj( O, p, j ); \\
\>\>\>\>        \keyw{if} ( T.find( q ) ) \comm{already}\\
\>\>\>\>\>          T.remove( q ); \comm{extracted}\\
\>\>\>\>        \keyw{else} \{ \\
\>\>\>\>\>          S.add( q );  L.push( q ); \\
\>\>\>\>\>          T.multipleInsert(q,dim()-2); \\
\> \} \>\> \} \> \} \\
\>  \keyw{return} S; \comm{T is empty at loop end} \\
\}
\end{algo}
\end{minipage}

\caption{Two digital hypersurface tracking algorithm: the Track (B)
algorithm requires an efficient ``is in set'' operation, the Track (C)
algorithm stores the list of cells that will be hit again by the
tracking. For the set T in (b), we have tried both \texttt{list} and
\texttt{multiset}. The former was much faster than the later in our
experiments.}
\label{fig:digital-surface-tracking}
\end{figure}

In the experiments, the object $O$ was a digital volumic
ball. Table~\ref{tbl:bdry-extraction} lists the running times
necessary to extract $\partial O$ for balls of various radii and
dimensions. The Scan (A) algorithm scans the whole image to find
boundaries. The Scan (B) algorithm scans the parallelepipedic subspace
containing the ball. The Track (A) algorithm extracts open or closed
boundaries from a starting bel (it follows both direct and indirect
bel adjacencies). Track (B) and (C) algorithms extract only closed
boundaries from a starting bel (they follow only direct bel
adjacencies).  All these algorithms are written generically and make
no assumption on the dimension of the image. The benchmarks show that
scanning algorithms depend on the size of the scanned subspace and
that tracking algorithms depend on the number of surfels in $\partial
O$. Running times are excellent since each bel is tracked in $\approx
1.7 \mu s$ in 3D (and $\approx 1.5 \mu s$ in 4D). Note that Track (B)
algorithm is much faster than Track (C) algorithm. This is because
\texttt{CharSet}s are efficient for the query ``is a cell in a given
set ?''.


\begin{table}
\caption{Running times for several boundary
extraction algorithms (see text).}
\label{tbl:bdry-extraction}
\begin{center}
\begin{tabular}{m{1cm}rrrm{1cm}m{1cm}m{1.4cm}m{1.4cm}m{1cm}}
Space size 
& Rad. 
& Nb spels & Nb surf. 
& Scan (A) 
& Scan (B) 
& Track (A) 
& Track (B) 
& Track (C) 
\\ \hline
$4096^2$ & 2000 & 12566345 & 16004 
& 2.07s & 2.00s & $<0.01$s & $<0.01$s & 0.01s
\\
$128^3$ & 30 & 113081 & 16926 
& 0.38s & 0.03s & 0.01s & 0.01s & 0.06s 
\\
$128^3$ & 60 & 904089 & 67734
& 0.39s & 0.34s & 0.07s & 0.05s & 0.57s%
\\
$256^3$ & 120 & 7236577 & 271350
& 3.15s & 2.70s & 0.36s & 0.32s & 5.24s 
\\
$512^3$ & 240 & 57902533 & 1085502
& 25.1s & 21.2s & 1.88s & 1.85s & 50.6s
\\
$64^4$ & 30 & 4000425 & 904648
& 4.26s & 4.00s & 1.91s & 1.37s & 4748s \\
\end{tabular}
\end{center}
\end{table}

\section{Conclusion}

We have presented a binary coding of every cell of the digital space
\CD{n}. This coding contains all the topological and geometric
information on the cell. It allows the design and implementation of
generic low-level algorithms that deals with subsets of
\CD{n}. Compact and efficient data structures can be built with this
coding. We illustrated the potential of this framework with a
classical digital topology application: boundary extraction. Arbitrary
dimensional algorithms are readily implemented in this framework and
benchmarks have proved that the resulting code is surprisingly
efficient in practice. Other digital topology and geometry
applications may be found in \cite{Lachaud02a}.



\bibliographystyle{plain}
\bibliography{cr-iwcia.bib}

\end{document}